\documentclass[12pt]{article}
\usepackage{amsmath}
\usepackage{amssymb}
\usepackage{amstext}
\usepackage{amsfonts}
\usepackage{graphicx,epsfig}
\usepackage{slashbox}
\usepackage{indentfirst}
\begin{document}

\baselineskip 20pt

\title{Study of Doubly Heavy Baryon Spectrum via QCD Sum Rules}
\author{ Liang Tang$^{a}$, Xu-Hao Yuan$^{a}$, Cong-Feng Qiao$^{a,b}$
 and Xue-Qian Li$^{a}$\\[0.5cm]
{\small $a)$ School of Physics, Nankai University, 300071, Tianjin,
China}\\
\small $b)$ Department of Physics, Graduate University, the Chinese
Academy of Sciences \\ \small YuQuan Road 19A, 100049, Beijing,
China}
\date{}
\maketitle

\begin{center}
\begin{minipage}{11cm}
In this work, we calculate the mass spectrum of doubly heavy baryons
with the diquark model in terms of the QCD sum rules. The
interpolating currents are composed of a heavy diquark field and a
light quark field. Contributions of the operators up to dimension
six are taken into account in the operator product expansion. Within
a reasonable error tolerance, our numerical results are compatible
with other theoretical predictions. This indicates that the diquark
picture reflects the reality and is applicable to the study of
doubly heavy baryons.

\end{minipage}
\end{center}

\section{Introduction}
The considerable success of quark model in interpreting a large
amount of hadronic observations has convinced people its undoubted
validity for many years. In the quark model, hadrons are constructed
according to two configurational schemes: mesons, consisting of a
quark and an antiquark ($q\bar{q}$); and baryons, consisting of
three quarks ($qqq$). Right after the birth of the quark model, the
diquark model was proposed where two quarks constitute a
color-anti-triplet which behaves as an independent object in the
baryon. In Gell-Mann's original paper on the quark model, he
discussed the possibility of the existence of free
diquarks\cite{GellMann:1964nj}. The concept of diquarks, has been
established in the fundamental theory, and has been invoked to help
illuminating a number of phenomena observed in experiments
\cite{Ida:1966ev,Lichtenberg:1975ap,Lichtenberg:1982jp,
Jaffe:2004ph,Wilczek:2004im,Ke:2007tg}. The systems composed of
three quarks should be described by the Faddev equations, but since
there are three coupled differential equations, solving them is
extremely difficult. As a matter of fact, the three-body problem is
still an unsolved subject even in classical physics. It is tempted
to consider the diquark-quark structure which turns the three-body
system into a two-body one, and the three Faddev equations then
reduce to single equation (no mater relativistic or
non-relativistic). Thus the problem is greatly simplified and
solution concerning baryon physics is obtained. However, for the
baryons which are composed of three light quarks, the three Faddev
equations have the same weight, so a problem emerges right away,
namely which two quarks are combined to compose a diquark while the
rest one moves independently. It seems to be an unbeatable
difficulty. However, recently the topic on diquarks revives, for it
may bring up some direct phenomenological consequences. Especially,
when there are two heavy quarks in a baryon, they may constitute a
relatively tight structure, a diquark. A diquark has the quantum
numbers of a two-quark system. For the ground state, a diquark has
positive parity and may be an axial-vector ($S=1$) or a scalar
($S=0$). According to the basic principle of QCD, for the two quarks
residing in a color anti-triplet, the interaction between them is
attractive.

Baryons containing two heavy quarks are important and intriguing
systems  to study the quark-diquark structure of baryons. The two
heavy quarks (b and c) can constitute a stable bound state of
$\bar{3}$, namely, a diquark which serves as a source of static
color field for light quarks \cite{Falk:1993gb}. The SELEX
Collaboration reported the first observation of a doubly charmed
baryon via the decay process $\Xi_{cc}^+\rightarrow\Lambda_c^+
K^-\pi^+$, by which its mass of 3519$\pm$ 1$\text{MeV}/c^2$
\cite{Mattson:2002vu} was determined. Later, this baryon was
confirmed by the SELEX Collaboration through the measurement of a
different decay mode $\Xi_{cc}^+\rightarrow p D^+K^-$, whose final
state involves a charmed meson \cite{Ocherashvili:2004hi}. However,
both the BABAR and Belle Collaborations did not observe this state
in $e^+e^-$ annihilation experiments
\cite{Aubert:2006qw,Chistov:2006zj}. This may be due to the distinct
beam structures of the two types of experiments, and the reason is
worthy of further and careful studies.

In the theoretical aspect, there have been numerous works in
studying the doubly heavy baryons
\cite{Majethiya:2008ia,Tong:1999qs, Ebert:2002ig,
He:2004px,Kiselev:2002iy}. All these works concern the dynamics
which results in the substantial diquark structure.

Although the QCD is proven to be an undisputablly valid theory about
strong interaction, the non-perturbative QCD which dominates the low
energy physics phenomena has not been fully understood yet. Among
the the theoretical methods in dealing with the non-perturbative
effects, the framework of the QCD Sum Rules which is indeed a bridge
between the short-distance and long-distance QCD as initiated by
Shifman {\it et al}. \cite{Shifman}, turns out to be a remarkably
successful and powerful technique for computing the hadronic
properties. Recently, a number of works have been worked out to
interpret the newly observed mesonic resonances within the framework
of the QCD sum rules
\cite{Wang:2009hiWang:2010uf,Zhang:2009em,Qiao:2010zh}. Meanwhile
with the QCD sum rules, a few works were performed in studying the
mass spectrum of doubly heavy baryons
\cite{Kiselev:2001fw,Bagan:1992za,
Zhang:2008rt,Albuquerque:2010bd,Narison:2010py}. In those studies,
the authors calculated the correlation function of baryonic currents
composed of quark fields by virtue of the operator product expansion
(OPE).

Since the correlation of the two heavy quarks is strong, they are
tempted to be bound into a diquark which can be regarded to manifest
independent degrees of freedom in the baryon. In this work, we  no
longer treat the two heavy quarks as independent constituents, but a
combined sub-system$-$diquark which behaves as a component of doubly
heavy baryons, and the corresponding field is denoted by a new
bosonic symbol $\Phi$ with a mass $m_D$. In fact, this picture was
recently proposed in Ref.\cite{Kim:2011ut}. Then, in this tentative
model for calculating the mass spectrum of doubly heavy baryon
systems, the light quark q (q=u,d,s) orbits the heavy diquark which
is a tightly bound QQ' (Q=c,b) pair. The application of the diquark
can simplify the interpolating currents which are important for
obtaining the baryon spectrum in the QCD sum rules. The spin-parity
quantum number of a ground-state diquark is either $0^+$ or $1^+$.
The former, along with a light q, can form the state with
$J^P=\frac{1}{2}^+$; the latter can form not only the state with
$J^P=\frac{1}{2}^+$, but also $J^P=\frac{3}{2}^+$. That is to say,
using the model of  the diquark and the QCD sum rules, we can study
the doubly heavy baryons with spin-parity $J^P=\frac{1}{2}^+$ and
$J^P=\frac{3}{2}^+$.

The content of the paper is arranged as follows. In Sec.II we derive
the formulas of the correlation function of the interpolating
currents with proper quantum numbers in terms of the QCD sum rules.
In Sec. III, our numerical results and relevant figures are
presented. Section IV is devoted to a summary and concluding
remarks.

\section{Formalism}
The method of the QCD Sum Rules is starting with choosing proper
correlation function (or Green's function) both at the quark-gluon
level and the hadron level. The correlation function for the doubly
heavy baryons reads
\begin{eqnarray}
\Pi(q^2)=i\int d^4xe^{iq\cdot x}\langle0| T\{J(x)\bar{J}(0)\}|0
\rangle\; .
\end{eqnarray}
Considering the spinor structures of baryons, the correlation
function has the Lorentz covariant expression as follows
\cite{Bagan:1992za,Zhang:2008rt}:
\begin{eqnarray}
\Pi(q^2)=\rlap /q \Pi_1(q^2)+\Pi_2(q^2)\; .
\end{eqnarray}
For each invariant function of $\Pi_1(q^2)$ and $\Pi_2(q^2)$ in the
doubly heavy baryons one can obtain a sum rule.

Following Refs.\cite{Bagan:1992za,Zhang:2008rt} and based on the
diquark model, the interpolating currents, which play a crucial role
in our analysis, are chosen to be
\begin{eqnarray}
J(x)&=&\Phi^a_\mu(x)\;\Gamma_k^{\mu}\;q^a(x), \hspace{2cm} {\rm
for\; spin\; 1/2\; baryon};\\
J^*_\mu(x)&=&\Phi^a_\mu(x)\;\Gamma_k\;q^a(x),\hspace{2cm} {\rm
for\; spin\; 3/2\; baryon};\\
J^\prime(x)&=&\Phi^a(x)\;\Gamma_k\;q^a(x), \hspace{2cm} {\rm for \;
spin\; 1/2\; baryon},
\end{eqnarray}
where $\Phi^a_\mu(x)$ and $\Phi^a(x)$ are axial vector and scalar
diquarks, respectively. The interpolating current $J$ corresponds to
$\Xi_{QQ^\prime}$ and $\Omega_{QQ^\prime}$, $J^*_\mu$ corresponds to
$\Xi^*_{QQ^\prime}$ and $\Omega^*_{QQ^\prime}$ and $J^\prime$
corresponds to $\Xi^\prime_{QQ^\prime}$ and
$\Omega^\prime_{QQ^\prime}$ respectively with $Q, Q^\prime=c,\;b$.
The concrete definition of $\Gamma_k^{\mu}$ and $\Gamma_k$ are
presented in Table \ref{diquark-Gamma}.

\begin{table}[h]
\begin{center}
\caption{The choice of $\Gamma_k^{\mu}$ and $\Gamma_k$. The index D
in $J_D^{P_D}$ means the diquark. $\Phi_{\{QQ^\prime\}}$ denotes the
axial vector diquark and $\Phi_{[QQ^\prime]}$ denotes the scalar
diquark respectively.}\label{diquark-Gamma}
\vspace{0.5cm}
\begin{tabular}{cccccc}
\hline\hline Baryon & Constituent  & $J^P$ & $J_D^{P_D}$ &
$\Gamma_k^{\mu}$ & $\Gamma_k$
\\
\hline $\Xi_{QQ^\prime}$ & $\Phi_{\{QQ^\prime\}}\;q$ & ${1\over2}^+$
& $1^+$ & $\gamma^\mu\gamma_5$ &-
\\
$\Xi^*_{QQ^\prime}$ & $\Phi_{\{QQ^\prime\}}\;q$ & ${3\over2}^+$
& $1^+$ & - & 1\\
$\Omega_{QQ^\prime}$ & $\Phi_{\{QQ^\prime\}}\;s$ & ${1\over2}^+
$ & $1^+$ & $\gamma^\mu\gamma_5$ &- \\
$\Omega^*_{QQ^\prime}$ & $\Phi_{\{QQ^\prime\}}\;s$ & ${3\over2}^+
$ & $1^+$ & -  & 1\\
$\Xi^\prime_{QQ^\prime}$ & $\Phi_{\left[QQ^\prime\right]}\;q$ &
${1\over2}^+$ & $0^+$ & - & 1\\
$\Omega^\prime_{QQ^\prime}$ & $\Phi_{\left[QQ^\prime\right]}\;s$ &
${1\over2}^+$ & $0^+$ &  -& 1
\\ \hline\hline
\end{tabular}%
\end{center}
\end{table}

On the phenomenological side, the correlation function is
expressed as a dispersion integral over a  physical regime,
\begin{eqnarray}
\Pi(q^2)=\lambda_H^2\frac{\rlap/ q+M_H}{M_H^2-q^2}+\int_{s_0}^\infty
ds \frac{\rho^h(s)} {s-q^2}+\cdots, \label{baryon}
\end{eqnarray}
where $M_H$ is the mass of the doubly heavy baryon, $\lambda_H$ is
the baryon coupling constant and $\rho^h(s)$ is the physical
spectral function of the continuum states. When we attain the above
expression, the summing relations for the Dirac and Rarita-Schwinger
spinors have been used, namely, for spin-3/2 baryons the numerator
of the first term in Eq.(\ref{baryon}) should be replaced by the
proper Lorentz structure \cite{Zhang:2008rt}
$$(\rlap/ q+M_H)(g_{\mu\nu}-\frac{1}{3}\gamma_\mu\gamma_\nu+\frac{
q_\mu\gamma_\nu-q_\nu\gamma_\mu}{3M_H}-\frac{2q_\mu
q_\nu}{3M_H^2})\; .$$

With the operator product expansion (OPE), the correlation
function $\Pi_i(q^2)$ (i=1 or 2) can be written as:
\begin{eqnarray}
\Pi_i(q^2)=\Pi_i^{\text{pert}}(q^2)+\sum_{\text{dim}=3}^{6}
\Pi_i^{\text{cond,\;dim}}(q^2)\; .
\end{eqnarray}
Here, ``pert", ``cond" and ``dim" refer to perturbative QCD
calculation, the quark or gluon condensates, and the relevant
condensate dimensions, respectively. $\Pi_i^{\text{pert}}(q^2)$ is
obtained by taking the absorptive part of the Feynman diagram A, and
$\Pi_i^{\text{cond,dim}}(q^2)$ represents the contributions from
various condensates. In this work, we consider the condensates up to
dimension six, as people usually do in the literature.

\begin{figure}
\begin{center}
\includegraphics[width=10cm]{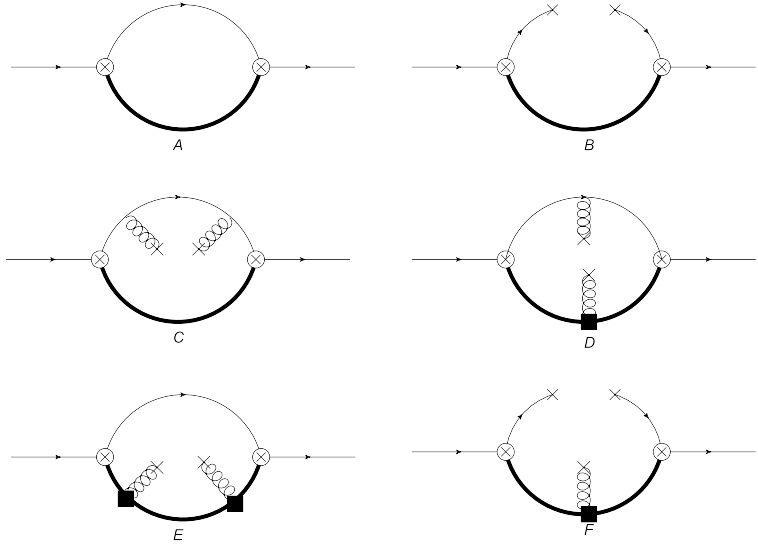}\\
\caption{Feynman diagrams.}\label{feynmandiagrams}
\end{center}
\end{figure}

\begin{figure}
\begin{center}
\includegraphics[width=8cm]{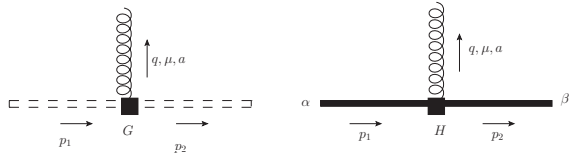}\\
\caption{Gluon-Diquark vertices. The wave line represents the gluon,
the dashed represents the scalar diquark, and  the black line
represents the axial vector diquark.}\label{vertex}
\end{center}
\end{figure}

The Feynman diagrams contributing to the correlation function of the
doubly heavy baryons are displayed in Fig.\ref{feynmandiagrams}, and
the gluon-diquark vertices are shown in Fig.\ref{vertex}. In
Ref.\cite{Jakob:1993th}, the effective gluon-diquark vertices were
given, which will also be used in our later calculations, so we just
copy them below:
\begin{eqnarray}
\text{SgS}&=&ig_st^a(p_1+p_2)_\mu F_S(Q^2),\\
\text{VgV}&=&-g_st^a\big{\{}g_{\alpha\beta}(p_1+p_2)_\mu-g_{\mu
\alpha}[(1+\kappa_V)p_1-\kappa_Vp_2]_\beta\nonumber\\
&-&g_{\mu\beta}[(1+\kappa_V)p_2-\kappa_Vp_1]_\alpha\big{\}}F_V(Q^2)\;
.
\end{eqnarray}
Here,\;$Q^2$=$-(p_1-p_2)^2$, $g_s=\sqrt{4\pi\alpha_s}$ denotes the
QCD coupling constant, $\kappa_v$ is the anomalous (chromo) magnetic
moment of the vector diquark and $t^a(=\lambda^a/2)$ is the
Gell-Mann color matrix. Furthermore, $F_S(Q^2)$ and $F_V(Q^2)$ are
the diquark form factors.

The scalar diquark's propagator is $\frac{i}{p^2-m_d^2+i\epsilon}$,
and the axial-vector diquark's propagator is
$\frac{-i\left(g_{\mu\nu}-{p_\mu p_\nu\over
m_d^2}\right)}{p^2-m_d^2+i\epsilon}$. Since the diquarks are made up
of two heavy quarks, according to the general rule, their
condensates are negligible \cite{Reinders:1984sr}.

The Feynman diagrams are computed not only with the regular QCD
Feynman rules, but also with the effective vertices displayed above
for point-like diquarks. For taking into account the composite
nature of diquarks, phenomenological form factors ($F_S(Q^2)$,
$F_V(Q^2)$ ) are introduced. The authors of Ref.\cite{Jakob:1993th}
suggested the vertex functions as following:
\begin{eqnarray}
F_S(Q^2)&=&\delta_S\frac{Q_S^2}{Q_S^2+Q^2}\;,\\
\delta_S&=&\alpha_s(Q^2)/\alpha_s(Q_S^2) \;\;\;\text{if}
\;\left(Q^2\geq Q_S^2\right)\;, \nonumber
\\\delta_S&=&1 \;\;\;\text{if}\;\left(Q^2<Q_S^2\right)\nonumber\;;\\
F_V(Q^2)&=&\delta_V(\frac{Q_V^2}{Q_V^2+Q^2})^2\;,\\
\delta_V&=&\alpha_s(Q^2)/\alpha_s(Q_V^2) \;\;\;\text{if}\;
\left(Q^2\geq Q_V^2\right) \nonumber\;,
\\\delta_V&=&1 \;\;\;\text{if}\;\left(Q^2<Q_V^2\right)\nonumber\;,
\end{eqnarray}
where $Q_S^2$=3.22$\text{GeV}^2$,\;$Q_V^2$=1.50$\text{GeV}^2$, and
the values of these special characteristic quantities are fixed by
fitting data.

Supposing the quark-hadron duality, the resultant sum rule for the
mass of the doubly heavy baryon reads
\begin{eqnarray}\label{massfunction}
m_H = \sqrt{-\frac{R_1}{R_0}},\label{mass}
\end{eqnarray}
with
\begin{eqnarray}
R_0 &=& \frac{1}{\pi}\int^{s_0}_{(m_{d}+m_q)^2}ds
\rho_i^{\text{pert}}(s)e^{-s/M_B^2} + \hat{\bf
B}[\Pi_i^{\text{cond,\;3}}(q^2)]
+\hat{\bf B}[\Pi_i^{\text{cond,\;4}}(q^2)]\nonumber\\
&&+\hat{\bf B}[\Pi_i^{\text{cond,\;5}}(q^2)]+\hat{\bf
B}[\Pi_i^{\text{cond,\;6}}(q^2)]\;,\label{R0}\\
R_1 & = &\frac{\partial}{\partial{M_B^{-2}}}{R_0}\;.
\end{eqnarray}
Here,$m_q$(q=u, d, or s) denotes the masses of the light quarks,
$m_d$ is the mass of the diquark, $M_B$ is the Borel parameter and
$s_0$ is the threshold cutoff introduced to remove the contribution
of the higher excited and continuum states \cite{P.Col}.

The perturbative contribution $\rho_1(s)$ and non-perturbative
contributions \\$\hat{\bf{B}}[\Pi_1^{\text{cond, dim}}(q^2)]$ for
$\Xi_{QQ^\prime}$ and $\Omega_{QQ^\prime}$ in Eq.(\ref{R0}) are
shown as follows:
\begin{eqnarray}
\rho_1(s)&=&-\frac{3 \left(m_d^2-m_q^2-s\right)
\sqrt{\left(m_d^2-m_q^2+s\right){}^2-4 s m_d^2}}{16\pi s^2}\;,
\end{eqnarray}
\begin{eqnarray}
\hat{\bf
B}[\Pi_1^{\text{cond,\;3}}(q^2)]&=&\langle\bar{\Psi}\Psi\rangle\frac{m_q
(3 M_B^4+m_d^2 m_q^2)}{12 M_B^6} e^{-\frac{m_d^2}{M_B^2}}\;,
\end{eqnarray}
\begin{subequations}\label{GG}
\begin{eqnarray}
\hat{\bf B}[\Pi_1^{\text{cond,\;4,\;C}}(q^2)]&=&\langle \alpha_s
G^2\rangle\int_0^1 dx
e^{-\frac{\frac{m_d^2}{1-x}+\frac{m_q^2}{x}}{M_B^2}} \bigg\{\frac{3
x^2-10 x+10}{192 \pi  M_B^2}+\frac{1}{192 \pi  (x-1)
x^2M_B^4}\nonumber\\
&&\times\Big[{x^2 \left(-3 x^2+10 x-10\right) m_d^2+(x-1)^2 \left(3
x^2-7 x+3\right) m_q^2}\Big] \nonumber\\
&&+\frac{1}{384 \pi  (x-1)^2 x^3 M_B^6}\Big[2 x^2 \left(-3 x^3+13
x^2-21 x+11\right) m_d^2 m_q^2\nonumber\\
&&+x^4 (3 x-10) m_d^4+(x-1)^3 (3 x^2-7 x+8 m_q^4\Big]+\frac{1}{384
\pi (x-1)^3 x^4 M_B^8}\nonumber\\
&&\Big[\left(\left(x^2-4 x+3\right) m_q^2-x^2 m_d^2\right)
\left((x-1)^2 m_q^2-x^2 m_d^2\right){}^2\Big]\bigg\}\;,\\
\hat{\bf B}[\Pi_1^{\text{cond,\;4,\;D}}(q^2)]&=&\langle \alpha_sG^2
\rangle\int_0^1 dx
e^{-\frac{\frac{m_d^2}{1-x}+\frac{m_q^2}{x}}{M_B^2}}
\bigg\{-\frac{1}{512 \pi (x-1) M_B^2}\Big[2 x^2 \kappa _v-x
\left(\kappa _v-2\right)-9 \kappa _v\nonumber\\
&&-6\Big]-\frac{1}{512 \pi (x-1)^2 x M_B^4}\Big[x^2 m_d^2 \left(-2 x
\kappa _v+\kappa _v-2\right)+(x-1)^2 m_q^2 (2 x \kappa
_v\nonumber\\
&&+\kappa _v+2)\Big]-\frac{\kappa _v \left((x-1)^2 m_q^2-x^2
m_d^2\right){}^2}{512 \pi (x-1)^3 x^2 M_B^6}\bigg\}\;,\\
 \hat{\bf
B}[\Pi_1^{\text{cond,\;4,\;E}}(q^2)]&=&\langle\alpha_sG^2\rangle\int_0^1dx
e^{-\frac{\frac{m_d^2}{1-x}+\frac{m_q^2}{x}}{M_B^2}}
\bigg\{\frac{1}{384 \pi  (x-1)^2 M_B^2}\Big[x (x^2 \left(6
\kappa _v+3\right)+x (4 \kappa _v^2-4 \kappa _v\nonumber\\
&&-2)+4 \kappa _v^2+6 \kappa _v+4)\Big]+\frac{1}{384 \pi (x-1)^3
M_B^4}\Big[(x-1)^2 m_q^2 (x (6 \kappa _v +3)\nonumber\\
&&+4 \kappa _v^2+2 \kappa _v+1)-x m_d^2 \left(x^2 \left(6 \kappa
_v+3\right)+x \left(4 \kappa _v^2-4 \kappa _v-2\right)+2 \kappa
_v\right)\Big]\nonumber\\
&&+\frac{1}{256 \pi (x-1)^3 x M_B^6}\Big[\left(2 \kappa _v+1\right)
(-x \left(2 x^2-3 x+1\right) m_d^2 m_q^2+x^3 m_d^4\nonumber\\
&&+(x-1)^3 m_q^4)\Big]\bigg\}\;,
\end{eqnarray}
\end{subequations}
\begin{eqnarray}
\hat{\bf B}[\Pi_1^{\text{cond,\;5}}(q^2)]&=&
\frac{g_s\langle\bar{\Psi}T\sigma\cdot G\Psi \rangle m_q
e^{-\frac{m_d^2}{M_B^2}}}{192
M_B^6}\left(3 M_B^2 \left(2 \kappa _v+1\right)-8 m_d^2\right)\;,\\
\hat{\bf B}[\Pi_1^{\text{cond,\;6}}(q^2)]
&=&\frac{g_s^2\langle\bar{\Psi}\Psi\rangle^2 e^{-\frac{m_d^2}
{M_B^2}}}{1296 M_B^6} \left(9 M_B^2 \left(2 \kappa _v+1\right)+17
m_d^2\right)\;,
\end{eqnarray}
where the superscripts (C, D and E) on the left hand of
Eq.(\ref{GG}) correspond to the labels in Fig.\ref{feynmandiagrams}.

For concision of the text, the detailed expressions of $R_0$ are
given in the Appendix.

\section{Numerical Analysis}
The numerical parameters used in this work are taken as
\cite{Narison:2010wb,Nakamura:2010zzi}
\begin{eqnarray}
\begin{aligned}
&\langle\bar{\Psi}\Psi\rangle=-(0.254\pm0.015 \text{GeV})^3,
& &\alpha_s\langle G^2\rangle=0.07\pm0.02\text{GeV}^4,\\
&g_s\langle \bar{\Psi}T\sigma\cdot G  \Psi\rangle=m_0^2\langle
\bar{\Psi}\Psi\rangle,& &\alpha_s\langle \bar{\Psi}\Psi\rangle^2
=(2.1\pm0.3)\times10^{-4}\text{GeV}^6,\\
&m_0^2=0.8\pm0.2\text{GeV}^2, & &m_s=0.14\text{GeV},\\
&m_u\simeq m_d=0.005\text{GeV}, & &m_c=1.27\text{GeV},\\
&m_b=4.19\text{GeV}, & &m_{\eta_c}=2.980\text{GeV},\\
&m_{J/\psi}=3.097\text{GeV}, & &m_{\Upsilon}=9.460\text{GeV}.
\end{aligned}
\end{eqnarray}
In our numerical analysis, we find that the effect of the two-gluon
condensate is tiny, i.e., if we shift the contributions of two-gluon
condensate a little, the final mass of the corresponding baryon
hardly changes. We choose it as in Ref\cite{Jakob:1993th} :
$\kappa_v=1.39$.

In practice, large uncertainty remains in the evaluation of baryon
spectrum due to the input constituent quark masses. The mass of the
baryon can be decomposed as $m_D+m_q+\Delta E$, where $m_D,\; m_q$
and $\Delta E$ are the masses of the heavy diquark, light-quark and
the binding energy respectively. The binding energy is calculable
within certain theoretical framework, whereas the quark, including
the diquark, masses are usually not definite input parameters. We
have to choose a reasonable strategy to determine the diquark
masses, which influence more on the heavy baryon spectrum than light
quark masses. It goes as follows.

There are few data on the doubly heavy baryons available, thus we
cannot directly extract all necessary information from experimental
data so far. Fortunately, the mass of $\Xi_{cc}$ has been measured.
Because of a simple symmetry argument, we know that the $cc$ must
constitute a spin-1 state, thus the $cc$ diquark in $\Xi_{cc}$ is an
axial vector state. With the measured $\Xi_{cc}$ mass as input, we
determine $m_{\{cc\}}=2.77$ GeV. Then we still need to fix the
masses of $bc$ and $bb$ diqurks. Generally the effective potential
between the two heavy quarks $Q,\; Q'$ includes the Coulomb and
linear confinement pieces. From the textbooks of the quantum
mechanics, we know the solutions of the Schr\"odinger equations with
the Coulomb potential proportional to ${\alpha_s\over r} $(the
solution is the Lauerre polynomial) or the linear potential
proportional to $\kappa r$ (the solution is the Airy  function). The
binding energy in the case where only the Coulomb piece exists is
proportional to the reduced mass $m$ of the $QQ'$ system, instead,
for the case where only the linear potential exists, the binding
energy is proportional to $({\kappa^2\over m})^{1/3}$ if $\kappa$ is
independent of $m$. Thus for the Schr\"odinger equation whose
potential includes both the Coulomb and linear confinement pieces,
the contributions from the two pieces compete and the dependence of
the binding energy on the reduced mass is uncertain. Then we would
like to invoke the data.

As discussed above, $m_{\{cc\}}$ and $\Delta E_{\{cc\}}$ can be
directly extracted from the data, then we will use those values for
the $cc$ diquark and some proposed rules to fix  $m_{\{bc\}}$,
$\Delta E_{\{bc\}}$, $m_{\{bb\}}$ and $\Delta E_{\{bb\}}$. Now let
us make a plausible comparison of the quantities about diquarks with
the corresponding mesons of the same flavor and spin structure. We
can have $\Delta E_{\{b\bar c\}}$ and $\Delta E_{\{b\bar b\}}$ from
the relations: $m_{\{b\bar c\}}=m_b+m_c+\Delta E_{\{b\bar c\}}$ and
$m_{\{b\bar b\}}=2m_b+\Delta E_{\{b\bar b\}}$. Thus $\Delta
E_{\{c\bar c\}}$ and $\Delta E_{\{b\bar b\}}$ are obtained as
$M_{J/\psi}-2m_c$ and $M_{\Upsilon(1S)}-2m_b$.

The diquarks  $cc$ or $bb$ are color-anti-triplet axial vector
states, instead the meson $J/\psi$ or $\Upsilon(1S)$ are
color-singlet vector states of $c\bar c$ and $b\bar b$. The
effective potentials between $QQ $ and $Q\bar Q$ only differ by a
color factor, so that we may have
$$\Delta E_{\{cc\}}:\Delta E_{\{c\bar c\}}=\Delta E_{\{bc\}}:
\Delta E_{\{b\bar c\}}=\Delta E_{\{bb\}}:\Delta E_{\{b\bar b\}},$$
where the dependence of the binding energies on color and
constituent masses may cancel. Then we obtain the binding energy for
the axial vector $bb$. Since so far there are no data on $B_c^*$
available yet, we cannot determine  $\Delta E_{\{b\bar
c\}}$ in the above scheme, but need to invoke another way. Since
$bc$ diquark is composed of $c$ and $b$ quarks, it is natural to
think that an interpolation between $cc$ and $bb$ diquarks would be
a good approximation for the $bc$ diquark, thus we write
$$\Delta E_{\{bc\}}={1\over 2}[\Delta E_{\{b b\}}+\Delta E_{\{cc\}}
].$$

Unlike the $bb$ and $cc$ diquarks, $bc$ diquark can be either an
axial vector or a scalar. Now let us determine the mass of the
scalar $bc$ diquark. The mass difference between spin-1 and spin-0
two-quark systems is due to the spin-spin interactions. Such
interaction is proportional to $1/(m_Qm_{Q'})$. Since there lack
enough data for $b\bar c$ mesons, let us first start with the
charmmonia which are well measured and then generalize to the $b\bar
c$ mesons.

The difference of the binding energies  of $J/\psi$ and $\eta_c$ is
due to the spin-spin interaction between $c$ and $\bar c$, and
besides a color factor related to the SU(3) Casimir factor which is
${4\over 3}$ for a color singlet and ${2\over 3}$ for a
color-anti-triplet, the case for the $cc$ diquark is the same. Thus
we may write
$$\Delta E_{\{c\bar c\}}-\Delta E_{[c\bar c]}=(M_{J/\psi}-M_{\eta_c})
.$$
Then using $$\Delta E_{\{c\bar c\}}:\Delta E_{[c\bar c]}=\Delta
E_{\{bc\}}:\Delta E_{[b c]},$$
we fix $\Delta E_{[b c]}$.

With above analysis, the diqaurk masses which will be adopted in the
following numerical computations are displayed as following:
\begin{eqnarray}
m_{\{cc\}}^{fit}=2.77\text{GeV},\;m_{[cb]}=5.73\text{GeV},\;
m_{\{cb\}}=5.80\text{GeV},\;m_{\{bb\}}=8.83\text{GeV}.
\end{eqnarray}

\begin{figure}
\centerline{\epsfysize=5truecm \epsfbox{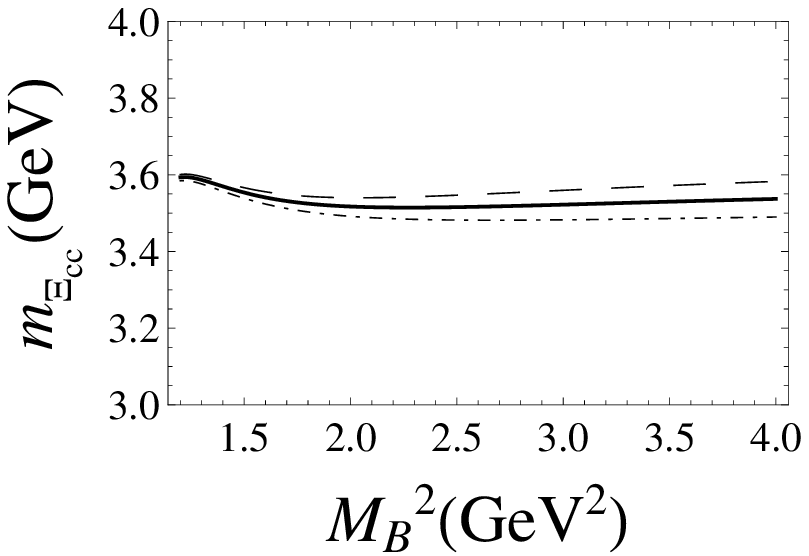}\epsfysize=5truecm
\epsfbox{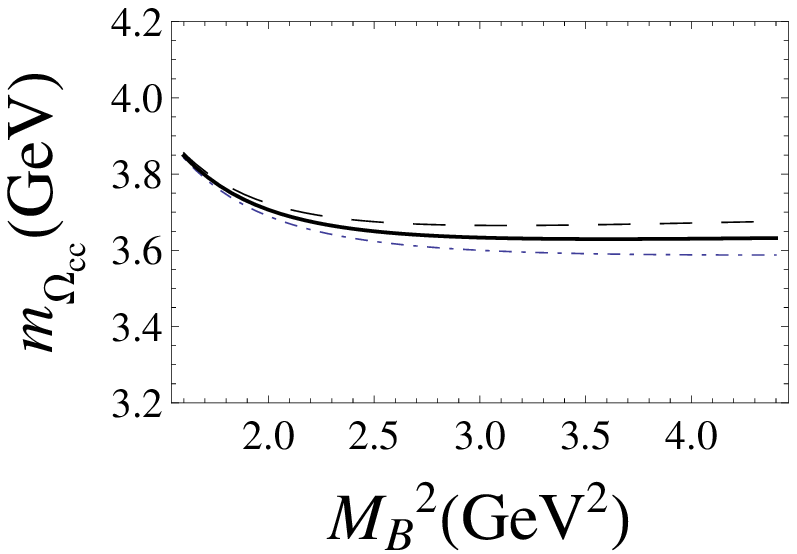}}\caption{Dependence of $\Xi_{cc}$ and
$\Omega_{cc}$ masses on the Borel parameter $M_B^2$. The continuum
thresholds $s_0$ are taken as $3.9^2, 4.0^2, 4.1^2 \text{GeV}^2$ for
$\Xi_{cc}$,  and $4.0^2, 4.1^2, 4.2^2 \text{GeV} ^2$ for
$\Omega_{cc}$, from down to up, respectively.We deliberately put two
vertical lines denoting the chosen Borel window.}
\label{xiccomegacc}
\centerline{\epsfysize=5truecm
\epsfbox{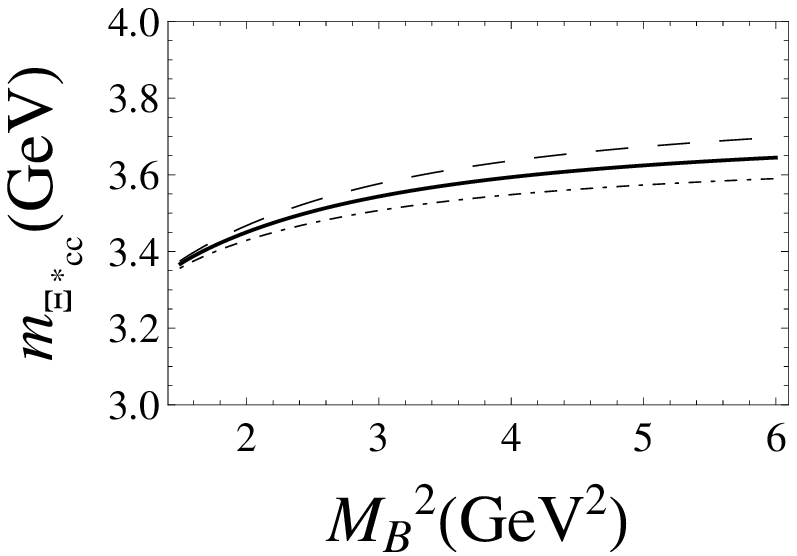}\epsfysize=5truecm
\epsfbox{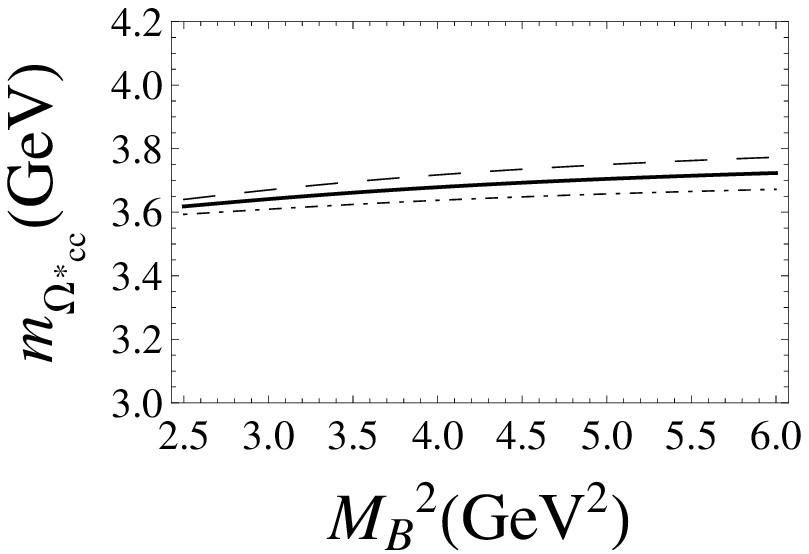}}\caption{Dependence of $\Xi^*_{cc}$ and
$\Omega^*_{cc}$ masses on Borel parameter $M_B^2$. The continuum
thresholds $s_0$ are taken as $4.0^2, 4.1^2, 4.2^2 \text{GeV}^2$ and
$4.1^2, 4.2^2, 4.3^2 \text{GeV}^2$, from down to up, respectively.We
deliberately put two vertical lines denoting the chosen Borel
window.}\label{xiccomegaccstar} \centerline{\epsfysize=5truecm
\epsfbox{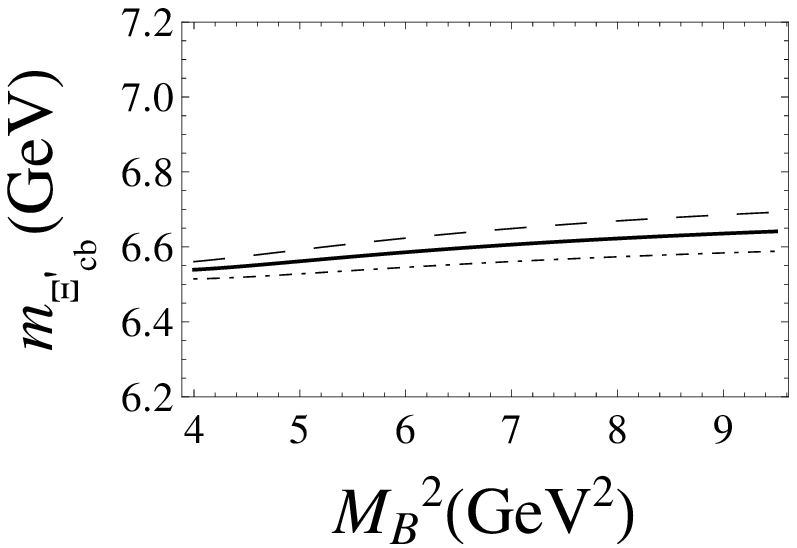}\epsfysize=5truecm
\epsfbox{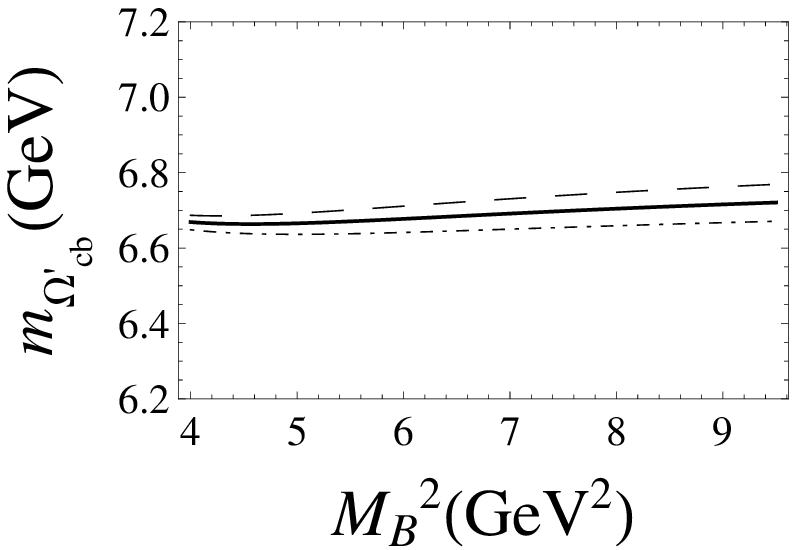}}\caption{Dependence of $\Xi^\prime_{cb}$
and $\Omega^\prime_{cb}$ masses on Borel parameter $M_B^2$. The
continuum thresholds $s_0$ are taken as $7.0^2, 7.1^2, 7.2^2
\text{GeV}^2$ and $7.1^2, 7.2^2, 7.3^2 \text{GeV}^2$, from down to
up, respectively.We deliberately put two vertical lines denoting the
chosen Borel window.}\label{xicbomegacbprime}
\end{figure}

\begin{figure}
\centerline{\epsfysize=5truecm \epsfbox{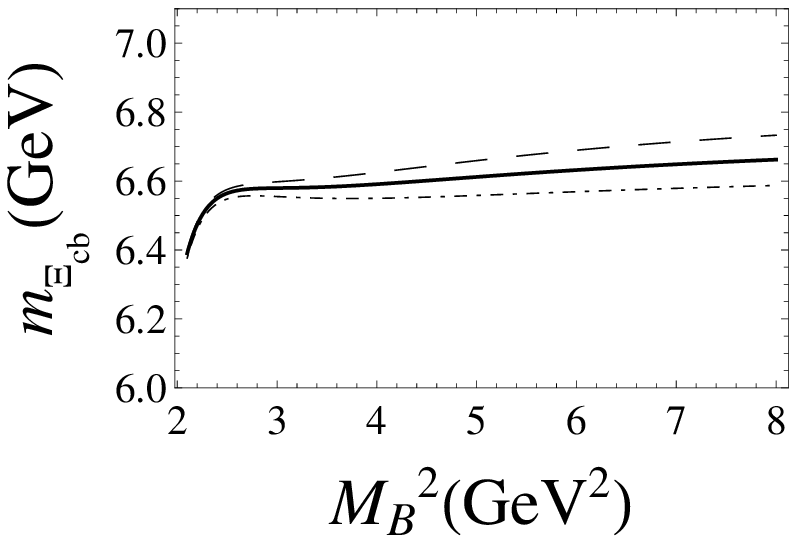}\epsfysize=5truecm
\epsfbox{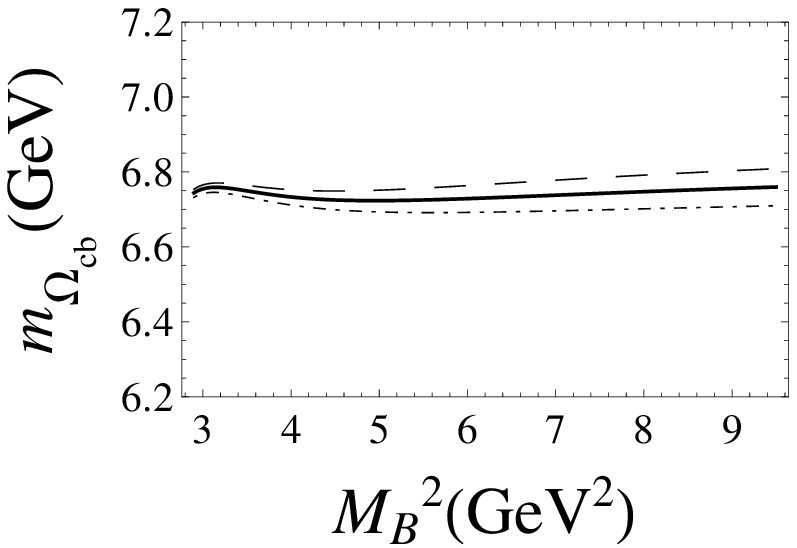}}\caption{Dependence of  $\Xi_{cb}$ and
$\Omega_{cb}$ masses on Borel parameter $M_B^2$. The continuum
thresholds $s_0$ are taken as $7.0^2, 7.1^2, 7.2^2  \text{GeV}^2$
and $7.1^2, 7.2^2, 7.3^2 \text{GeV} ^2$, from down to up,
respectively.We deliberately put two vertical lines denoting the
chosen Borel window.}\label{xicbomegacb}
\centerline{\epsfysize=5truecm
\epsfbox{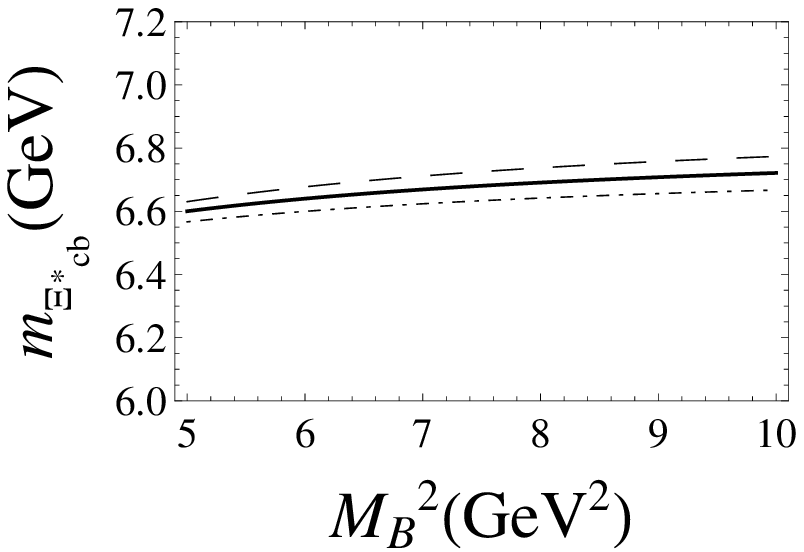}\epsfysize=5truecm
\epsfbox{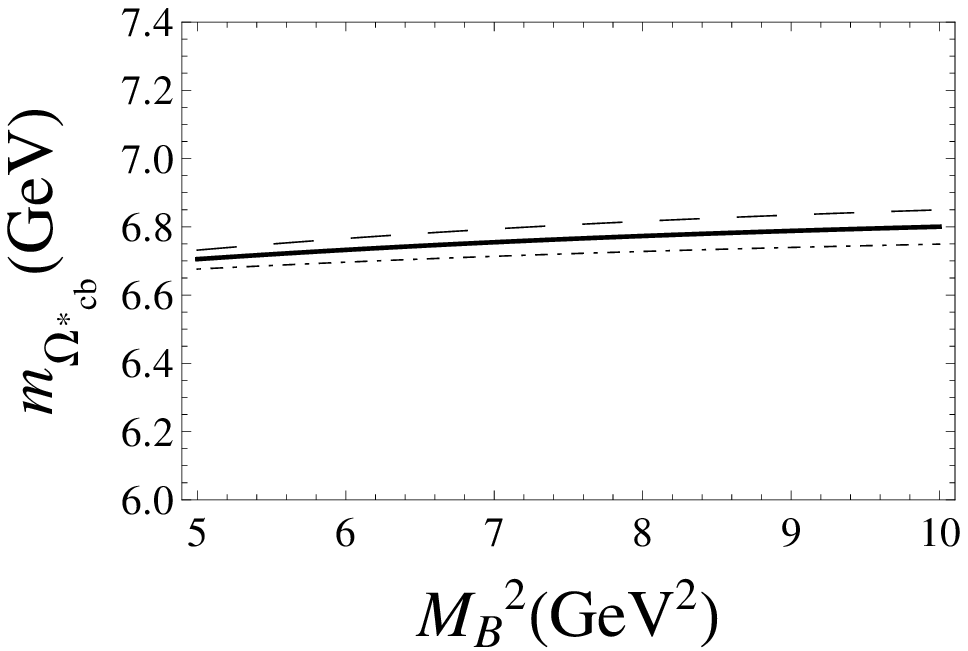}}\caption{Dependence of $\Xi_{cb}^*$ and
$\Omega_{cb}^*$ masses on Borel parameter $M_B^2$. The continuum
thresholds $s_0$ are taken as $7.1^2, 7.2^2, 7.3^2 \text{GeV}^2$ and
$7.2^2, 7.3^2, 7.4^2 \text{GeV} ^2$, from down  to up,
respectively.We deliberately put two vertical lines denoting the
chosen Borel window.}\label{xicbomegacbstar}
\centerline{\epsfysize=5truecm \epsfbox{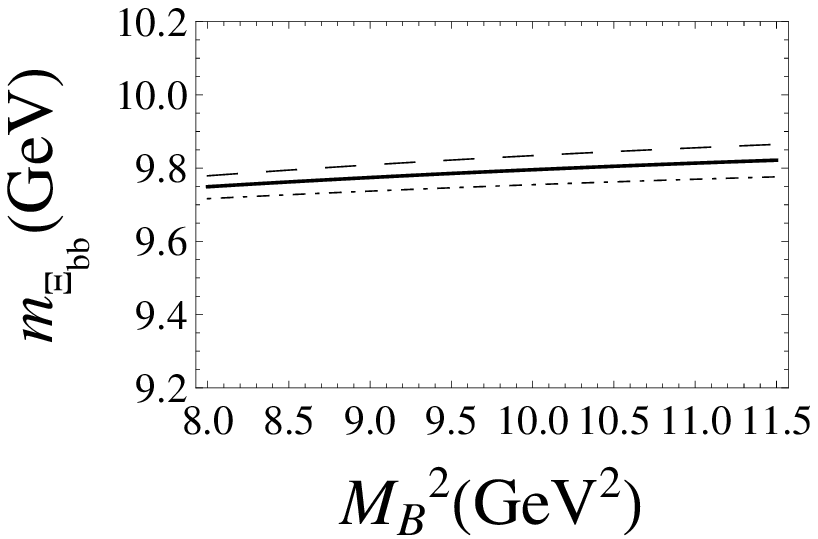}\epsfysize=5truecm
\epsfbox{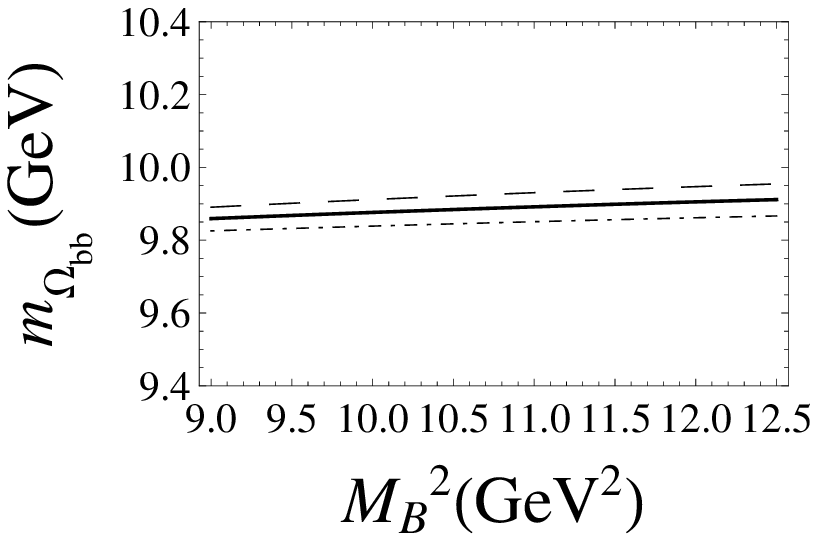}}\caption{Dependence of  $\Xi_{bb}$ and
$\Omega_{bb}$ masses on Borel parameter $M_B^2$. The continuum
thresholds $s_0$ are taken as $10.7^2, 10.8^2, 10.9^2 \text{GeV}^2$
and $10.9^2, 11.0^2, 11.1^2 \text {GeV}^2$, from down to
up,respectively.We deliberately put two vertical lines denoting the
chosen Borel window.}\label{xibbomegabb}
\end{figure}

\begin{figure}
\centerline{\epsfysize=5truecm
\epsfbox{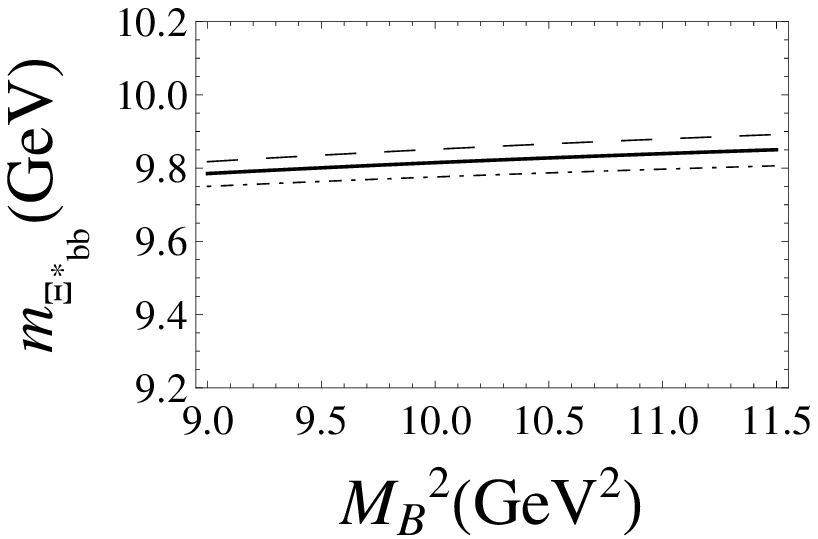}\epsfysize=5truecm
\epsfbox{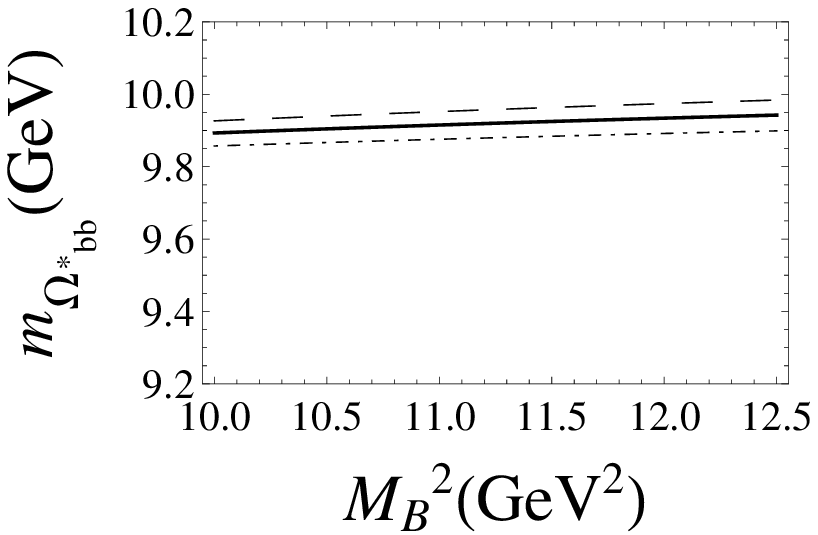}}\caption{Dependence of $\Xi_{bb}^*$ and
$\Omega_{bb}^*$ masses on Borel parameter $M_B^2$. The continuum
thresholds $s_0$ are taken as $10.4^2, 10.5^2, 10.6^2 \text{GeV}^2$
and $10.5^2, 10.6^2, 10.7^2 \text{GeV}^2$, from down  to up,
respectively. We deliberately put two vertical lines denoting the
chosen Borel window.}\label{xibbomegabbstar}
\end{figure}

For choosing the proper threshold $s_0$ and the Borel parameter
$M_B^2$ there are two criteria. First, the perturbative contribution
should be larger than the contributions from  all kinds of
condensates, and another is that the pole contribution should be
larger than the continuum contribution\cite{Shifman,Reinders:1984sr,
D.S.Du}. On the other hand, the dependence of the evaluated masses
of the doubly heavy baryons is rather unsensitive to variations of
the Borel parameter in the Borel windows. For each baryon
fortunately we can find an optimal Borel window where the two
aforementioned criteria are satisfied and the results are almost
independent of the Borel parameter after all. By the windows we
obtain the masses of doubly heavy baryons. The dependence are shown
in Figs.(\ref{xiccomegacc}-\ref{xibbomegabbstar}), respectively. The
numerical results are collected in Table \ref{mass-spectra} for
various quantum numbers. For a comparison with other theoretical
estimates on the baryon masses given in the literature, we also show
those results in Table \ref{comparison}. The error bars are
estimated by varying the Borel parameters, $s^0$ and the
uncertainties in the condensates as well. It is noted that the
uncertainty caused by introducing the diquark configuration is
included in the diquark form factors (Eqs(10) and (11)).

\begin{table}
\centering \caption{The mass spectra of doubly heavy baryons. The
``pole" stands for the contribution from the pole term to the
spectral density. The ``cond" stands for the contribution from the
condensate terms in the operator product expansion, where the
threshold parameter $s_0$ takes its central value. $\Delta m$ is the
energy gap between masses of other species of baryons and
$\Xi_{QQ'}$  with the same  diquark flavors.}\label{mass-spectra}
\vspace{0.3cm}
\renewcommand\arraystretch{1.4}
\begin{tabular}{|c|c|c|c|c|c|c|c|}
\hline\hline \small{Baryon} & \small{quark} & $J^P(J_D^{P_D})$   &
\small{Results(GeV)} &$\Delta m$ &$M_B^2(\text{GeV}^2)$
  &\small{pole}  & \small{cond}
\\
\hline $\Xi_{cc}$ & $\{cc\}q$ & ${1\over2}^+(1^+)$  &$3.519^{fit}$&
0 &2.0-3.8  &(53-84)\% &(3-18)\%
\\
\hline$\Omega_{cc}$ & $\{cc\}s$ & ${1\over2}^+(1^+)$   &
$3.63^{+0.06}_{-0.03}+\delta_{cc}$& 0.11 &2.2-4.0 &(58-83)\%
&(5-28)\%
\\
\hline$\Xi^*_{cc}$ & $\{cc\}q$ & ${3\over2}^+(1^+)$   &
$3.62^{+0.08}_{-0.09}+\delta_{cc}$&0.10&2.4-5.4  &(51-88)\% &(2-5)\%
\\
\hline$\Omega^*_{cc}$ & $\{cc\}s$ & ${3\over2}^+(1^+)$   &
$3.71^{+0.07}_{-0.05}+\delta_{cc}$&0.19 &3.0-5.5 & (55-82)\%&(3-8)\%
\\
\hline$\Xi^\prime_{cb}$ & $[cb]q$ & ${1\over2}^+(0^+)$   &
$6.61^{+0.08}_{-0.10}+\delta^1_{cb}$&0&5.0-8.0  & (54-79)\%&(2-6)\%
\\
\hline$\Omega^\prime_{cb}$ & $[cb]s$ & ${1\over2}^+(0^+)$  &
$6.69\pm0.06+\delta^1_{cb}$&0.08&5.0-8.0  &(58-81)\% &(3-11)\%
\\
\hline$\Xi_{cb}$ & $\{cb\}q$ & ${1\over2}^+(1^+)$ &
$6.65^{+0.07}_{-0.08}+\delta^2_{cb}$&0&3.5-7.0  &(58-90)\% &(3-21)\%
\\
\hline$\Omega_{cb}$ & $\{cb\}s$ & ${1\over2}^+(1^+)$   &
$6.75^{+0.05}_{-0.03}+\delta^2_{cb}$&0.10&4.0-8.0  &(55-87)\%
&(4-25)\%
\\
\hline$\Xi^*_{cb}$ & $\{cb\}q$ & ${3\over2}^+(1^+)$   &
$6.69\pm0.08+\delta^2_{cb}$&0.04& 5.2-9.0 &(50-79)\% &(1-2)\%
\\
\hline$\Omega^*_{cb}$ & $\{cb\}s$ & ${3\over2}^+(1^+)$   &
$6.77^{+0.06}_{-0.04}+\delta^2_{cb}$&0.12&6.0-9.0 &(54-75)\%
&(2-4)\%
\\
\hline$\Xi_{bb}$ & $\{bb\}q$ & ${1\over2}^+(1^+)$   &
$9.80\pm0.07+\delta_{bb}$&0&8.5-11.0 &(63-77)\% &(2-4)\%
\\
\hline$\Omega_{bb}$ & $\{bb\}s$ & ${1\over2}^+(1^+)$   &
$9.89^{+0.04}_{-0.03}+\delta_{bb}$&0.09& 9.5-12.0 &(73-84)\%
&(2-4)\%
\\
\hline$\Xi^*_{bb}$ & $\{bb\}q$ & ${3\over2}^+(1^+)$   &
$9.84\pm0.07+\delta_{bb}$&0.04&9.5-11.0  &(68-76)\% &1\%
\\
\hline$\Omega^*_{bb}$ & $\{bb\}s$ & ${3\over2}^+(1^+)$   &
$9.93^{+0.05}_{-0.04}+\delta_{bb}$&0.13&10.5-12.0  &(67-74)\% &2\%
\\ \hline\hline
\end{tabular}%
\end{table}
\begin{table}
\centering \caption{Comparison with other theoretical resutls and
the experimental data (if available). The quantities are in
GeV.}\label{comparison} \vspace{0.3cm}
\renewcommand\arraystretch{1.4}
\begin{tabular}{|c|c|c|c|c|c|c|c|}
\hline\hline Baryon & quark & $J^P(J_D^{P_D})$  & Our work
&\cite{Zhang:2008rt}&\cite{Ebert:2002ig} &\cite{He:2004px}
&Exp.\cite{Mattson:2002vu}
\\
\hline $\Xi_{cc}$ & $\{cc\}q$ & ${1\over2}^+(1^+)$ &$3.519^{fit}$
&4.26&3.620&3.520&3.519$\pm$0.001
\\
\hline$\Omega_{cc}$ & $\{cc\}s$ & ${1\over2}^+(1^+)$  & $3.63$
&4.25&3.778&3.619&-
\\
\hline$\Xi^*_{cc}$ & $\{cc\}q$ & ${3\over2}^+(1^+)$  &
$3.62$&3.90&3.727&3.630&-
\\
\hline$\Omega^*_{cc}$ & $\{cc\}s$ & ${3\over2}^+(1^+)$  &
$3.71$&3.81&3.872&3.721&-
\\
\hline$\Xi^\prime_{cb}$ & $[cb]q$ & ${1\over2}^+(0^+)$  &
$6.61$&6.95&6.963&7.028&-
\\
\hline$\Omega^\prime_{cb}$ & $[cb]s$ & ${1\over2}^+(0^+)$  &
$6.69$&7.02&7.116&7.116&-
\\
\hline$\Xi_{cb}$ & $\{cb\}q$ & ${1\over2}^+(1^+)$  &
$6.65$&6.75&6.933&6.838&-
\\
\hline$\Omega_{cb}$ & $\{cb\}s$ & ${1\over2}^+(1^+)$  &
$6.75$&7.02&7.088&6.941&-
\\
\hline$\Xi^*_{cb}$ & $\{cb\}q$ & ${3\over2}^+(1^+)$  &
$6.69$&8.00&6.980&6.986&-
\\
\hline$\Omega^*_{cb}$ & $\{cb\}s$ & ${3\over2}^+(1^+)$  &
$6.77$&7.54&7.130&7.077&-
\\
\hline$\Xi_{bb}$ & $\{bb\}q$ & ${1\over2}^+(1^+)$  &
$9.80$&9.78&10.202&10.272&-
\\
\hline$\Omega_{bb}$ & $\{bb\}s$ & ${1\over2}^+(1^+)$  &
$9.89$&9.85&10.359&10.369&-
\\
\hline$\Xi^*_{bb}$ & $\{bb\}q$ & ${3\over2}^+(1^+)$  &
$9.84$&10.35&10.237&10.337&-
\\
\hline$\Omega^*_{bb}$ & $\{bb\}s$ & ${3\over2}^+(1^+)$  &
$9.93$&10.28&10.389&10.429&-
\\ \hline\hline
\end{tabular}%
\end{table}

Note that inside the Tables(\ref{mass-spectra},\ref{comparison}),
the mass of the baryon $\Xi_{cc}$ with superscript ``fit" is taken
as inputs to obtain the mass of diquark $m_{\{cc\}}$ and then the
masses of other baryon states in the table are predicted.

As indicated in above the choice of the diquark masses is based on
our postulate about the binding energy, so this strategy would
certainly bring up some theoretical uncertainties. To explicitly
show how the diquark mass influences the spectrum of doubly heavy
baryons, let us shift the corresponding diquark masses by 0.1 GeV,
and we find that the uncertainty of the baryon mass lies within
$0.047\sim 0.064 \;$GeV. One can be convinced that the uncertainty
should be no more than 10\% as we change the diquark mass within a
reasonable range. Moreover, in the forth column of
Table\ref{mass-spectra} we put a term $\delta_{QQ'}$ following the
predicted mass to manifest a possible error. In next column of this
table, we list the gaps ($\Delta m$) among the concerned baryon
masses where the uncertainties cancel out, and hence may make more
senses. That means the predictions in this work, especially on the
mass gaps, are experimentally testable.

\section{Summary and Conclusions}
In this work, the masses of various doubly heavy baryons have been
studied in terms of the QCD sum rules where the diquark structures
are priori assumed. In the calculation we keep the contributions of
the condensates up to dimension six in OPE. Our results, in certain
tolerance, are in accordance with the theoretical predictions via
other models. Especially, it is worth pointing out that our results
are reasonably consistent with that calculated in the QCD sum rules
without assuming diquark structures.

In the calculation, an effective coupling between diquark and gluon
which was phenomenologically introduced is adopted. The form factor
at the effective vertex indeed manifests an inner structure of the
diquark. But as the diquark is viewed as an independent degree of
freedom, this factor performs as an ad-hoc parameter in the given
theory and it plays a role just as the quark or gluon condensates in
the QCD sum rules which were obtained either from an underlying
theory (such as the value of the gluon condensate could be obtained
from the dilute gas approximation of instantons) or by fitting data
(such as the value of the quark condensate might be gained by
fitting the pion decay constant).

Our results imply that the structure of a heavy diquark and a light
quark is indeed a reasonable configuration for the doubly heavy
baryons. The Large Hadron Collider (LHC) which has already begun
running, even at lower energy (7 TeV) and luminosity , will provide
a large database of doubly heavy baryons. Once enough data are
available, one can further analyze the doubly heavy baryons of
various flavors and spins. Comparing our theoretical predictions on
their mass spectra with the data, will not only enrich our knowledge
on the underlying theory, i.e. the low energy QCD, but also further
investigate the diquark structure and applicability for dealing with
the processes such as production and decay of the doubly heavy
baryons.

\vspace{.7cm} {\bf Acknowledgments} \vspace{.3cm}

This work was supported in part by the National Natural Science
Foundation of China(NSFC) and by the CAS Key Projects KJCX2-yw-N29
and H92A0200S2.


\appendix{\bf\Large Appendix}

The perturbative contribution $\rho_2(s)$ and nonperturbative
contributions $\hat{\bf{B}}[\Pi_2^{\text{cond,\;dim}}(q^2)]$ for
$\Xi_{QQ^\prime}$ and $\Omega_{QQ^\prime}$ in Eq.(\ref{R0}) are
shown as follows:
\begin{eqnarray*}
&&\rho_2(s)=-\frac{3 m_q \left(m_d^2-m_q^2-s\right)
\sqrt{\left(m_d^2-m_q^2+s\right){}^2-4 s m_d^2}}{8 \pi
s^2}\;,\\
&&\hat{\bf B}[\Pi_2^{\text{cond,\;3}}(q^2)]=-\frac{ (2 M_B^4 +m_d^2
m_q^2)}{2 M_B^4}\langle\bar{\Psi}\Psi\rangle e^{-\frac{m_d^2}
{M_B^2}}\;,
\end{eqnarray*}
\begin{eqnarray*}
\hat{\bf
B}[\Pi_2^{\text{cond,\;4,\;C}}(q^2)]&=&\langle\alpha_sG^2\rangle
\int_0^1dxe^{-\frac{\frac{m_d^2}{1-x}+\frac{m_q^2}{x}}{M_B^2}}
\bigg\{\frac{ m_q}{16 \pi  x^2 M_B^2}-\frac{1}{16 \pi (x-1) x^3
M_B^4}\\
&&\times\Big[m_q \left((x-1)^2
m_q^2-x^2 m_d^2\right)\Big]\\
&&+\frac{1}{96 \pi (x-1)^2 x^4 M_B^6}\\
&&\times\Big[m_q \left(-4 (x-1) x^3 m_d^2 m_q^2+x^4 m_d^4+(x-1)^3 (3
x-1) m_q^4\right)\Big]\bigg\}\;,
\end{eqnarray*}
\begin{eqnarray*}
\hat{\bf
B}[\Pi_2^{\text{cond,\;4,\;D}}(q^2)]&=&\langle\alpha_sG^2\rangle
\int_0^1dxe^{-\frac{\frac{m_d^2}{1-x}+\frac{m_q^2}{x}}{M_B^2}}
\bigg\{\frac{1}{64 \pi (x-1) x M_B^2}\Big[m_q \left(2 \kappa
_v+1\right)\Big]\\
&&+\frac{1}{256 \pi (x-1)^2 x^2 M_B^4}\Big[m_q \left(x^2 m_d^2
\left(\kappa _v+2\right)-(x-1)^2 m_q^2 \left(3 \kappa
_v+2\right)\right)\Big]\\
&&-\frac{1}{256 \pi (x-1)^3 x^3 M_B^6}\Big[m_q \kappa _v
\left((x-1)^2 m_q^2-x^2 m_d^2\right){}^2\Big]\bigg\}\;,
\end{eqnarray*}
\begin{eqnarray*}
\hat{\bf
B}[\Pi_2^{\text{cond,\;4,\;E}}(q^2)]&=&\langle\alpha_sG^2\rangle
\int_0^1dxe^{-\frac{\frac{m_d^2}{1-x}+\frac{m_q^2}{x}}{M_B^2}}
\bigg\{\frac{1}{192 \pi
(x-1)^2 M_B^2}\Big[m_q \left(8 \kappa _v^2+8 \kappa _v+5\right)\Big]\\
&&+\frac{1}{192 \pi  (x-1)^3 x M_B^4}\Big[m_q \left(2 (x-1)^2 m_q^2
\left(2 \kappa _v^2+2 \kappa _v-1\right)\right.\\
&&\left.-x m_d^2 \left(x \left(4 \kappa _v^2+4 \kappa
_v-2\right)+3\right)\right)\Big] -\frac{1}{128 \pi (x-1)^2 x^2
M_B^6}\\
&&\times\Big[m_q \left((x-1) m_q^2-x m_d^2\right){}^2\Big]\bigg\}\;,
\end{eqnarray*}
\begin{eqnarray*}
\hat{\bf B}[\Pi_2^{\text{cond,\;5}}(q^2)]&=&-
\frac{g_s\langle\bar{\Psi}T\sigma\cdot G\Psi\rangle
e^{-\frac{m_d^2}{M_B^2}}}{32 M_B^4}
\left(M_B^2 \left(2 \kappa _v+1\right)-8 m_d^2\right),\\
\hat{\bf B}[\Pi_2^{\text{cond,\;6}}(q^2)]&=&0.
\end{eqnarray*}

The perturbative contributions $\rho_i(s)$ and nonperturbative
contributions $\hat{\bf{B}}[\Pi_i^{\text{cond, dim}}(q^2)]$ for
$\Xi^*_{QQ^\prime}$ and $\Omega^*_{QQ^\prime}$ in Eq.(\ref{R0}) are
shown as follows:
\begin{eqnarray*}
&&\rho_1(s)=-\frac{3 (-m_d^2+m_q^2+s)
\sqrt{(m_d^2-m_q^2+s){}^2-4 s m_d^2}}{8 \pi  s^2},\\
&&\rho_2(s)=-\frac{3 m_q \sqrt{(m_d^2-m_q^2+s){}^2-4 s m_d^2}}{4 \pi
s},\\
&&\hat{\bf B}[\Pi_1^{\text{cond,\;3}}(q^2)]=-\frac{m_q (3
M_B^4+m_d^2 m_q^2)}{6 M_B^6}\langle\bar{\Psi}
\Psi\rangle e^{-\frac{m_d^2}{M_B^2}},\\
&&\hat{\bf B}[\Pi_2^{\text{cond,3}}(q^2)]= \frac{ (2 M_B^4+m_d^2
m_q^2)}{2 M_B^4}\langle\bar{\Psi}\Psi\rangle
e^{-\frac{m_d^2}{M_B^2}},
\end{eqnarray*}
\begin{eqnarray*}
\hat{\bf B}[\Pi_1^{\text{cond,\;4,\;C}}(q^2)]&=&
\langle\alpha_sG^2\rangle\int_0^1dx
e^{-\frac{\frac{m_d^2}{1-x}+\frac{m_q^2}{x}}{M_B^2}}
\bigg\{\frac{x-2}{96 \pi M_B^2}+\frac{1}{192 \pi  (x-1)^2 x^3
M_B^4}\Big[2 (x-1)^4 x m_q^2\\
&&-2 x^3 \left(x^2-3 x+2\right) m_d^2\Big]+\frac{1}{192 \pi (x-1)^2
x^3 M_B^6}\Big[-2 (x-2) (x-1) x^2 m_d^2 m_q^2\\
&&+x^4 m_d^4+(x-3) (x-1)^3 m_q^4\Big]\bigg\},\\
\hat{\bf
B}[\Pi_2^{\text{cond,\;4,\;C}}(q^2)]&=&\langle\alpha_sG^2\rangle\int_0^1dx
e^{-\frac{\frac{m_d^2}{1-x}+\frac{m_q^2}{x}}{M_B^2}} \bigg\{-\frac{
m_q}{48 \pi x^2 M_B^2}\\
&&+\frac{1}{96 \pi (x-1) x^3 M_B^4}\Big[m_q \left(2 (x-1)^2
m_q^2-x^2 m_d^2\right)\Big]\\
&&-\frac{1}{96 \pi (x-1) x^4 M_B^6}\Big[m_q^3 \left((x-1)^2
m_q^2-x^2 m_d^2\right)\Big]\bigg\}\;,
\end{eqnarray*}
\begin{eqnarray*}
\hat{\bf B}[\Pi_1^{\text{cond,\;4,\;D}}(q^2)]&=&0\;,\\
\hat{\bf B}[\Pi_2^{\text{cond,\;4,\;D}}(q^2)]&=&0\;,
\end{eqnarray*}
\begin{eqnarray*}
\hat{\bf
B}[\Pi_1^{\text{cond,\;4,\;E}}(q^2)]&=&\langle\alpha_sG^2\rangle\int_0^1
dx e^{-\frac{\frac{m_d^2}{1-x}+\frac{m_q^2}{x}}{M_B^2}}
\bigg\{\frac{1}{192 \pi (x-1)^3 M_B^2}\Big[x (-x^2 \left(4
\kappa _v^2+4 \kappa _v+7\right)\\
&&+4 \left(\kappa _v^2+\kappa _v+1\right)+3 x^3)\Big]+\frac{1}{192
\pi  (x-1)^3 M_B^4}\Big[x^2 m_d^2 (4 \left(\kappa _v^2+\kappa
_v+1\right)\\
&&-3 x)+(x-1)^2 m_q^2 \left(3 x-\left(2 \kappa
_v+1\right){}^2\right)\Big]+\frac{1}{128 \pi (x-1)^2 x
M_B^6}\\
&&\times\Big[{\left((x-1) m_q^2-x m_d^2\right){}^2}\Big]\bigg\}\;,\\
\hat{\bf
B}[\Pi_2^{\text{cond,\;4,\;E}}(q^2)]&=&\langle\alpha_sG^2\rangle\int_0^1
dxe^{-\frac{\frac{m_d^2}{1-x}+\frac{m_q^2}{x}}{M_B^2}}
\bigg\{-\frac{1}{192 \pi (x-1)^2 M_B^2}\Big[m_q \left(8 \kappa
_v^2+8 \kappa _v+5\right)\Big]\\
&&+\frac{1}{192 \pi (x-1)^3 x M_B^4}\\
&&\times\Big[m_q \left(x m_d^2 \left(x \left(4 \kappa _v^2+4 \kappa
_v-2\right)+3\right)-2 (x-1)^2 m_q^2 \left(2 \kappa _v^2+2 \kappa
_v-1\right)\right)\Big]\\
&&+\frac{1}{128 \pi (x-1)^2 x^2 M_B^6}\Big[m_q \left((x-1) m_q^2-x
m_d^2\right){}^2\Big]\bigg\}\;,
\end{eqnarray*}
\begin{eqnarray*}
&&\hat{\bf B}[\Pi_1^{\text{cond,\;5}}(q^2)]
=\frac{g_s\langle\bar{\Psi}T \sigma\cdot G\Psi\rangle m_d^2 m_q
e^{-\frac{m_d^2}{M_B^2}}}
{12 M_B^6}\;,\\
&&\hat{\bf B}[\Pi_2^{\text{cond,\;5}}(q^2)]
=-\frac{g_s\langle\bar{\Psi}T\sigma\cdot G\Psi\rangle m_d^2
e^{-\frac{m_d^2}{M_B^2}}}{4 M_B^4}\;,\\
&&\hat{\bf B}[\Pi_1^{\text{cond,\;6}}(q^2)]
=-\frac{g_s^2\langle\bar{\Psi}\Psi\rangle^2  m_d^2 e^{-\frac
{m_d^2}{M_B^2}}}{81 M_B^6}\;,\\
&&\hat{\bf B}[\Pi_2^{\text{cond,\;6}}(q^2)]=0\;.
\end{eqnarray*}

The perturbative contributions $\rho_i(s)$ and nonperturbative
contributions $\hat{\bf{B}}[\Pi_i^{\text{cond, dim}}(q^2)]$ for
$\Xi^\prime_{QQ^\prime}$ and $\Omega^\prime_{QQ^\prime}$ in
Eq.(\ref{R0}) are shown as follows:
\begin{eqnarray*}
&&\rho_1(s)=\frac{3 \left(-m_d^2+m_q^2+s\right)\sqrt{\left(m_d^2-
m_q^2+s\right){}^2-4 s m_d^2}}{32 \pi  s^2},\\
&&\rho_2(s)=\frac{3 m_q \sqrt{\left(m_d^2-m_q^2+s\right){}^2-4 s
m_d^2}}{16 \pi s},\\
&&\hat{\bf B}[\Pi_1^{\text{cond,\;3}}(q^2)]=\frac{m_q  \left(3 M_B^4
+m_d^2 m_q^2\right)}{24 M_B^6}\langle\bar{\Psi}\Psi\rangle
e^{-\frac{m_d^2}{M_B^2}},\\
&&\hat{\bf B}[\Pi_2^{\text{cond,\;3}}(q^2)]=-\frac{ \left(2 M_B^4+
m_d^2 m_q^2\right)}{8 M_B^4}\langle \bar{\Psi}\Psi\rangle
e^{-\frac{m_d^2}{M_B^2}}\;,
\end{eqnarray*}
\begin{eqnarray*}
\hat{\bf
B}[\Pi_1^{\text{cond,\;4,\;C}}(q^2)]&=&\langle\alpha_sG^2\rangle\int_0^1d
x e^{-\frac{\frac{m_d^2}{1-x}+\frac{m_q^2}{x}}{M_B^2}}
\bigg\{\frac{3 x^2-10 x+10}{384 \pi M_B^2}+\frac{1}{768 \pi (x-1)^3
x^4 M_B^4}\\
&&\times\Big[{2 (x-1)^4 x^2 \left(3 x^2-7 x+3\right) m_q^2-2 (x-1)^2
x^4 \left(3 x^2-10 x+10\right) m_d^2}\Big]\\
&&+\frac{1}{768 \pi (x-1)^3 x^4 M_B^6}\Big[2 (x-1) \left(-3 x^3+13
x^2-21 x+11\right) x^3 m_d^2 m_q^2\\
&&+\left(3 x^2-13 x+10\right) x^5 m_d^4+(x-1)^3 \left(3 x^3-10
x^2+15 x-8\right) x m_q^4\Big]\\
&&+\frac{1}{768 \pi (x-1)^3 x^4 M_B^8}\Big[(x-1) x^4 (3 x-5) m_d^4
m_q^2\\
&&+(7-3 x) (x-1)^3 x^2 m_d^2 m_q^4-x^6 m_d^6+(x-3) (x-1)^5
m_q^6\Big]\bigg\},
\end{eqnarray*}
\begin{eqnarray*}
\hat{\bf
B}[\Pi_2^{\text{cond,\;4,\;C}}(q^2)]&=&\langle\alpha_sG^2\rangle\int_0^1d
xe^{-\frac{\frac{m_d^2}{1-x}+\frac{m_q^2}{x}}{M_B^2}} \bigg\{\frac{
m_q}{64 \pi  x^2 M_B^2}-\frac{1}{64 \pi  (x-1) x^3 M_B^4}\\
&&\times\Big[m_q \left((x-1)^2 m_q^2-x^2
m_d^2\right)\Big]+\frac{1}{384 \pi (x-1)^2 x^4 M_B^6}\\
&&\times\Big[m_q \left(-4 (x-1) x^3 m_d^2 m_q^2+x^4 m_d^4+(x-1)^3 (3
x-1) m_q^4\right)\Big]\\
&&-\frac{1}{384 \pi (x-1)^2 x^5 M_B^8}\Big[m_q^3 \left((x-1)^2
m_q^2-x^2 m_d^2\right){}^2\Big]\bigg\}\;,
\end{eqnarray*}
\begin{eqnarray*}
&&\hat{\bf B}[\Pi_1^{\text{cond,\;4,\;D}}(q^2)]=0\;,\\
&&\hat{\bf B}[\Pi_2^{\text{cond,\;4,\;D}}(q^2)]=0\;,\\
&&\hat{\bf B}[\Pi_1^{\text{cond,\;4,\;E}}(q^2)]=0\;,\\
&&\hat{\bf B}[\Pi_2^{\text{cond,\;4,\;E}}(q^2)]=0\;,\\
&&\hat{\bf
B}[\Pi_1^{\text{cond,\;5}}(q^2)]=\frac{g_s\langle\bar{\Psi}
T\sigma\cdot G\Psi\rangle m_d^2 e^{-\frac{m_d^2}{M_B^2}}}
{16 M_B^4}\;,\\
&&\hat{\bf B}[\Pi_2^{\text{cond,\;5}}(q^2)]=-\frac{g_s\langle\bar
{\Psi}T\sigma\cdot G\Psi\rangle m_d^2 m_q e^{-\frac{m_d^2}{M_B^2}}}
{48 M_B^6}\;,\\
&&\hat{\bf B}[\Pi_1^{\text{cond,\;6}}(q^2)]=\frac{g_s^2\langle\bar
{\Psi}\Psi\rangle^2 m_d^2 e^{-\frac{m_d^2}{M_B^2}}}{324 M_B^6}\;,\\
&&\hat{\bf B}[\Pi_2^{\text{cond,\;6}}(q^2)]=0\;.
\end{eqnarray*}


\begin{thebibliography}{99}
\bibitem{GellMann:1964nj}
M.~Gell-Mann, Phys.\ Lett.\  {\bf 8}, 214 (1964).

\bibitem{Ida:1966ev}
M.~Ida and R.~Kobayashi, Prog.\ Theor.\ Phys.\  {\bf 36} (1966) 846.

\bibitem{Lichtenberg:1975ap}
D.~B.~Lichtenberg, Nuovo Cim.\  A {\bf 28}, 563 (1975).

\bibitem{Lichtenberg:1982jp}
D.~B.~Lichtenberg, W.~Namgung, E.~Predazzi and J.~G.~Wills, Phys.\
Rev.\ Lett.\  {\bf 48}, 1653 (1982).

\bibitem{Jaffe:2004ph}
R.~L.~Jaffe, Phys.\ Rept.\  {\bf 409}, 1 (2005) [Nucl.\ Phys.\
Proc.\ Suppl.\  {\bf 142}, 343 (2005)].

\bibitem{Wilczek:2004im}
F.~Wilczek, arXiv:hep-ph/0409168.

\bibitem{Ke:2007tg}
H.~W.~Ke, X.~Q.~Li and Z.~T.~Wei, Phys.\ Rev.\  D {\bf 77}, 014020
(2008) [arXiv:0710.1927 [hep-ph]].

\bibitem{Falk:1993gb}
A.~F.~Falk, M.~E.~Luke, M.~J.~Savage and M.~B.~Wise, Phys.\ Rev.\  D
{\bf 49}, 555 (1994) [arXiv:hep-ph/9305315].

\bibitem{Mattson:2002vu}
M.~Mattson {\it et al.}  [SELEX Collaboration], Phys.\ Rev.\ Lett.\
{\bf 89}, 112001 (2002) [arXiv:hep-ex/0208014].

\bibitem{Ocherashvili:2004hi}
A.~Ocherashvili {\it et al.}  [SELEX Collaboration], Phys.\ Lett.\ B
{\bf 628}, 18 (2005) [arXiv:hep-ex/0406033].

\bibitem{Aubert:2006qw}
B.~Aubert {\it et al.}  [BABAR Collaboration], Phys.\ Rev.\  D {\bf
74}, 011103 (2006).

\bibitem{Chistov:2006zj}
R.~Chistov {\it et al.}  [BELLE Collaboration], Phys.\ Rev.\ Lett.\
{\bf 97}, 162001 (2006).

\bibitem{Majethiya:2008ia}
A.~Majethiya, B.~Patel, A.~K.~Rai and P.~C.~Vinodkumar,
arXiv:0809.4910 [hep-ph].

\bibitem{Tong:1999qs}
S.~P.~Tong, Y.~B.~Ding, X.~H.~Guo, H.~Y.~Jin, X.~Q.~Li, P.~N.~Shen
and R.~Zhang, Phys.\ Rev.\  D {\bf 62}, 054024 (2000)
[arXiv:hep-ph/9910259].

\bibitem{Ebert:2002ig}
D.~Ebert, R.~N.~Faustov, V.~O.~Galkin and A.~P.~Martynenko, Phys.\
Rev.\  D {\bf 66}, 014008 (2002) [arXiv:hep-ph/0201217].

\bibitem{He:2004px}
D.~H.~He, K.~Qian, Y.~B.~Ding, X.~Q.~Li and P.~N.~Shen, Phys.\ Rev.\
D {\bf 70}, 094004 (2004) [arXiv:hep-ph/0403301].

\bibitem{Kiselev:2002iy}
V.~V.~Kiselev, A.~K.~Likhoded, O.~N.~Pakhomova, V.~A.~Saleev, Phys.\
Rev.\  {\bf D66}, 034030 (2002). [hep-ph/0206140].

\bibitem{Shifman}
M. A. Shifman, A. I. Vainshtein and V. I. Zakharov, Nucl. Phys. {\bf
B147}, 385 (1979); ibid, Nucl. Phys. {\bf B147}, 448 (1979).

\bibitem{Wang:2009hiWang:2010uf}
Z.~G.~Wang and X.~H.~Zhang, Commun.\ Theor.\ Phys.\  {\bf 54}, 323
(2010) [arXiv:0905.3784 [hep-ph]]; Z.~G.~Wang, Y.~M.~Xu and
H.~J.~Wang, Commun.\ Theor.\ Phys.\  {\bf 55}, 1049 (2011)
[arXiv:1004.0484 [hep-ph]].

\bibitem{Zhang:2009em}
J.~R.~Zhang and M.~Q.~Huang, Commun.\ Theor.\ Phys.\  {\bf 54}, 1075
(2010) [arXiv:0905.4672 [hep-ph]].

\bibitem{Qiao:2010zh}
C.~F.~Qiao, L.~Tang, G.~Hao and X.~Q.~Li, J.\ Phys.\ G {\bf 39},
015005 (2012) arXiv:1012.2614 [hep-ph].

\bibitem{Kiselev:2001fw}
V.~V.~Kiselev, A.~K.~Likhoded, Phys.\ Usp.\  {\bf 45}, 455-506
(2002) [hep-ph/0103169].

\bibitem{Bagan:1992za}
E.~Bagan, M.~Chabab and S.~Narison, Phys.\ Lett.\  B {\bf 306}, 350
(1993).

\bibitem{Zhang:2008rt}
J.~R.~Zhang and M.~Q.~Huang, Phys.\ Rev.\  D {\bf 78}, 094007 (2008)
[arXiv:0810.5396 [hep-ph]].

\bibitem{Albuquerque:2010bd}
R.~M.~Albuquerque and S.~Narison, Nucl.\ Phys.\ Proc.\ Suppl.\  {\bf
207-208}, 265 (2010) [arXiv:1009.2428 [hep-ph]].

\bibitem{Narison:2010py}
S.~Narison and R.~Albuquerque, Phys.\ Lett.\  B {\bf 694}, 217
(2010) [arXiv:1006.2091 [hep-ph]].

\bibitem{Kim:2011ut}
K.~Kim, D.~Jido and S.~H.~Lee, arXiv:1103.0826 [nucl-th].

\bibitem{Jakob:1993th}
R.~Jakob, P.~Kroll, M.~Schurmann and W.~Schweiger, Z.\ Phys.\  A
{\bf 347}, 109 (1993) [arXiv:hep-ph/9310227].

\bibitem{Reinders:1984sr}
L.~J.~Reinders, H.~Rubinstein and S.~Yazaki, Phys.\ Rept.\  {\bf
127}, 1 (1985).

\bibitem{P.Col}
P. Colangelo and A. Khodjamirian, in {\it At the frontier of
particle physics / Handbook of QCD}, edited by M. Shifman (World
Scientific, Singapore, 2001), arXiv:hep-ph/0010175.

\bibitem{Narison:2010wb}
S.~Narison, Nucl.\ Phys.\ Proc.\ Suppl.\  {\bf 207-208}, 315 (2010)
[arXiv:1010.1959 [hep-ph]].

\bibitem{Nakamura:2010zzi}
K.~Nakamura {\it et al.}  [Particle Data Group], J.\ Phys.\ G {\bf
37}, 075021 (2010).

\bibitem{D.S.Du}
D. S. Du, J. W. Li and M. Z. Yang, Phys. Lett. {\bf B619}, 105
(2005).
\end{thebibliography}
\end{document}